\documentclass[aps,pra,preprint,groupedaddress,showpacs,amsmath,letterpaper]{revtex4}
\usepackage{graphicx}

\begin{document}
\title{A two reservoir model of quantum error correction}
\author{James P. Clemens}
\affiliation{Department of Physics, University of Arkansas, Fayetteville, AR 72701}
\affiliation{Department of Physics, Miami University, Oxford, OH 45056}
\author{Julio Gea-Banacloche}
\affiliation{Department of Physics, University of Arkansas, Fayetteville, AR 72701}
\date{\today}                                           

\begin{abstract}
We consider a two reservoir model of quantum error correction with a hot bath causing errors in the qubits and a cold bath cooling the ancilla qubits to a fiducial state.  We consider error correction protocols both with and without measurement of the ancilla state.  The error correction acts as a kind of refrigeration process to maintain the data qubits in a low entropy state by periodically moving the entropy to the ancilla qubits and then to the cold reservoir.  We quantify the performance of the error correction as a function of the reservoir temperatures and cooling rate by means of the fidelity and the residual entropy of the data qubits.  We also make a comparison with the continuous quantum error correction model of Sarovar and Milburn [Phys.~Rev.~A {\bf 72} 012306].
\end{abstract}


\maketitle
\section{Introduction}
It is clear by now that the operation of a large scale quantum computer will require some form of protection of the data qubits from decoherence arising from interaction with their environment and imperfect control fields which are applied to carry out quantum logic gates.  There are three methods known to provide such protection --- quantum error correcting codes \cite{Laflamme-1996-198, Steane-1996-793, Calderbank-1996-1098, Shor-1995-R2493, Steane-2003-042322, DiVincenzo-1996-3260,Calderbank-1997-405, Knill-2000-2525}, decoherence free subspaces \cite{Zanardi-1997-3306, Lidar-1999-4556, Lidar-1998-2594, Bacon-2000-1758, Bacon-1999-1944}, and dynamical suppression of decoherence \cite{Viola-1998-2733}.

In this paper we focus on the role of ancilla preparation in a quantum error correction protocol.  We consider a situation where the ancilla qubits are reset to a fiducial state by coupling them to a cold reservoir.  We model the source of errors as a hot reservoir coupling to both the data and ancilla qubits.  This system can be viewed as a kind of refrigerator where the data qubits are preserved in a low entropy state via the error correction protocol with the excess entropy being dumped into the cold reservoir when the ancilla qubits are reset.  The performance of the code then depends critically upon the rate at which this cycle can be performed.  The resetting of the ancilla qubits also requires an energy input of at least $k_B T \log 2$ per bit of information erased where $T$ is the temperature of the environment into which the information is lost \cite{Landauer-1961-183, Piechocinska-2000-062314}.  This defines a minimum energy cost for running the error correction protocol which is different from that considered by one of the authors in Ref.~\cite{Gea-Banacloche-2002-217901}.

In a real system it may not be practical to control the coupling of the ancilla qubits to a cold reservoir at will.  It may be necessary to use an active cooling method which could be similar to that analyzed in \cite{Kosloff-2000-8093}.  Nevertheless we model the cooling of the ancilla qubits as a coupling to a thermal reservoir in this paper.

The paper is laid out as follows.  The physical system is described in Section~\ref{sec:model} along with the master equation describing the coupling to the reservoirs.  The error correction protocol with measurement is analyzed in Sec.~\ref{sec:measure}.  In Sec.~\ref{sec:nomeasure} we describe error correction without measurement and make a comparison with the continuous error correction described in \cite{Sarovar-2005-012306} followed by a discussion and conclusions in Sec.~\ref{sec:discuss}.

\section{\label{sec:model} The physical system}
We model a physical implementation of the three-bit repetition code protecting against bit-flip errors consisting of six individually addressable qubits, perhaps contained in an ion trap or array of ion microtraps.  For simplicity we assume that the logical $|0\rangle$ and $|1\rangle$ states are degenerate and coherent transitions are achieved by two-photon Raman pulses.

The accumulation of bit-flip errors in the qubits is modeled by a coupling to a hot thermal reservoir.  In practice errors arise due to a noisy environment and imperfect control fields applied to the qubits.  Here we simply replace these specific sources of errors with a coupling to a thermal reservoir with an effective temperature.  The important parameter is the heating rate $\gamma_h$ at which each qubit accumulates errors.  For qubits with degenerate logical states $|0\rangle$ and $|1\rangle$ such a coupling gives rise to a completely mixed steady state.  Quantum error correction slows the rate at which errors accumulate.

The interaction Hamiltonian for the system is
\begin{equation}
H_I = \sum_{i=1}^6\left\{f_i(t)\sigma_{ix} +  g_i(t)\sigma_{iz} + h_i(t)H_i + \sum_{j>i} k_{ij}(t)P_{ij}\right\}
\end{equation}
where $\sigma_{i\alpha}$ is a Pauli pseudo-spin operator acting on the $i$th qubit, $H_i$ is the Hadamard operator acting on the $i$th qubit, and $P_{ij}$ is the two-qubit phase shift operator
\begin{equation}
P_{ij} = {\rm diag}\{\alpha_{00}, \alpha_{01}, \alpha_{10}, \alpha_{11}\}
\end{equation}
represented in the computational subspace of qubits $i$ and $j$.  This type of operator can be realized for example by applying a state dependent force on a pair of ions in separate microtraps \cite{Cirac-2000-579, Calarco-2001-062304, Sasura-2003-062318}.  We make the simplifying assumption that the pushing gate can be carried out between any pair of qubits without extra swap gates.  This requires the ability to physically shuffle the qubits into proximity as needed.  The parameters $f_i(t), g_i(t), h_i(t)$, and $k_i(t)$ represent applied fields which can be turned on and off at will.

In addition to the unitary evolution the qubits are also coupled to two external thermal reservoirs at temperatures $T_h$ and $T_c$.  The reservoir at $T_h$ heats the qubits causing bit flips at a rate $\gamma_h$ while the reservoir at $T_c$ cools the ancilla qubits at a rate $\Gamma_c$.  We assume that the cooling can be turned on and off at will either by lifting the degeneracy of the logical states via an external field or by an active cooling method similar to that described in \cite{Kosloff-2000-8093}.  The system evolution is determined by the master equation
\begin{eqnarray}
\dot\rho &=& -\frac{i}{\hbar}\left[H_I(t),\rho\right] + \frac{\gamma_h}{2}\sum_{i=1}^6\left(\sigma_{ix}\rho\sigma_{ix} - 2\rho\right) + p(t)\frac{\Gamma_c}{2}(n_c+1)\sum_{i=4}^6\left(\sigma_{i-}\rho\sigma_{i+} - \sigma_{i+}\sigma_{i-}\rho - \rho\sigma_{i+}\sigma_{i-}\right) \nonumber \\ && + p(t)\frac{\Gamma_c}{2}n_c\sum_{i=4}^6\left(\sigma_{i+}\rho\sigma_{i-} - \sigma_{i-}\sigma_{i+}\rho - \rho\sigma_{i-}\sigma_{i+}\right)
\end{eqnarray}
where $p(t)$ is an additional control parameter which allows a variable coupling to the cold reservoir and $n_c = \left(\exp(\hbar\omega/k_BT_c)-1\right)^{-1}$ is the thermal photon number of the cold reservoir at the energy separating the $|0\rangle$ and $|1\rangle$ states of the qubits.  Note that this assumes the qubits are at least temporarily made nondegenerate.  For an active cooling scenario the parameter $n_c$ characterizes the residual probability to find an ancilla qubit in the $|1\rangle$ state following the cooling step.

\section{Error correction with measurement\label{sec:measure}}

\begin{figure}
   \centering
   \includegraphics{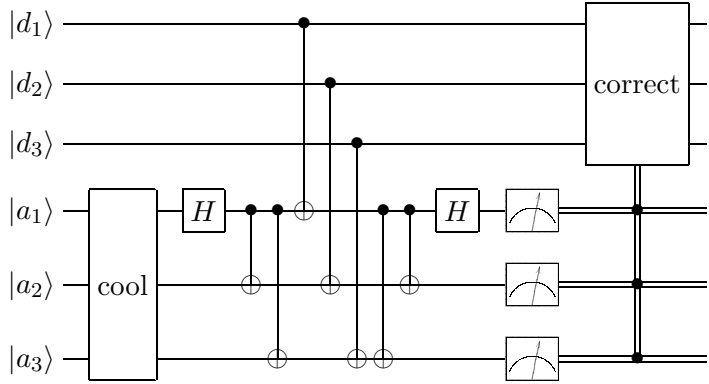} 
   \caption{Quantum circuit for one round of error correction.}
   \label{fig:protocol}
\end{figure}

The basic quantum error correction protocol is illustrated in Fig.~\ref{fig:protocol}.  It consists of a cooling step, ancilla preparation, transversal CNOT between the data qubits and ancilla qubits, ancilla decoding, syndrome measurement, and appropriate correction.  When this is successful it corrects bit-flip errors of weight 1 on the data qubits.

\begin{figure}[b]
   \centering
   \includegraphics{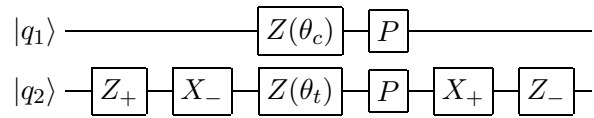} 
   \caption{Quantum circuit for a CNOT gate.  Here $\theta_c = (\alpha_{00}-\alpha_{10})$ and $\theta_t = (\alpha_{00}-\alpha_{01})$. The parameters $\alpha_{ij}$ are the diagonal elements of the pushing gate $P$.  $Z(\theta)$ is a rotation about the $z$-axis by an angle $\theta$ and likewise for $X(\theta)$.  $Z_\pm = Z(\pm\pi/2)$ and likewise for $X_\pm$.}
   \label{fig:expandcnot}
\end{figure}

\begin{figure}
   \centering
   \includegraphics[width=6in]{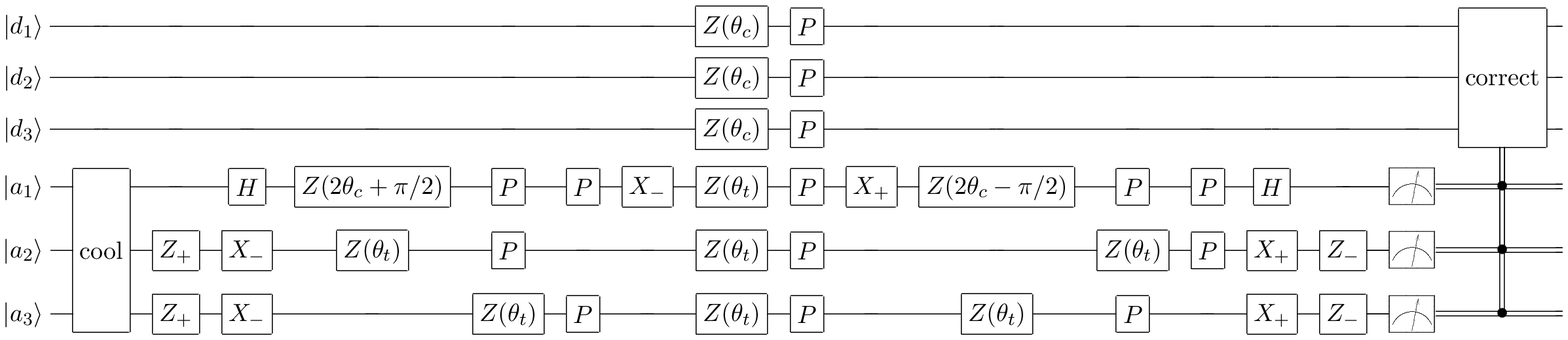} 
   \caption{Quantum circuit for one round of error correction expressed in terms of operation available in the Hamiltonian.  The gates and parameters are as defined in Fig.~\ref{fig:expandcnot}}
   \label{fig:expandprot}
\end{figure}

The fundamental gates available in the Hamiltonian do not include the CNOT gate.  Therefore we express the CNOT as a sequence of fundamental gates as shown in Fig.~\ref{fig:expandcnot}.  This leads to the full error correction circuit in Fig.~\ref{fig:expandprot}.

The error correction protocol in Fig.~\ref{fig:expandprot} is simulated by the method of quantum trajectories \cite{Carmichael-1993-book}.  In this method the master equation is unraveled as a sum over pure states
\begin{equation}
\rho(t) = \sum_{\rm REC} P_{\rm REC} |\psi_{REC}(t)\rangle\langle\psi_{\rm REC}(t)|
\end{equation}
conditioned on a measurement record $\rm REC$.  The system undergoes a continuous evolution governed by a non-Hermitian Hamiltonian
\begin{equation}
H_{\rm eff}(t) = H_I(t) - i\left(\frac{\gamma_h}{2} + p(t)\frac{\Gamma_c}{2}(n_c+1)\sum_{i=4}^6\sigma_{i+}\sigma_{i-} + p(t)\frac{\Gamma_c}{2}n_c\sum_{i=4}^6 \sigma_{i-}\sigma_{i+}\right)
\end{equation}
with discontinous bit-flip errors generated by the action of $\sigma_{ix}$ occurring with probability $\gamma_h dt$ on all the qubits, heating events generated by $\sigma_{i+}$ occurring with probability $p(t)\Gamma_c n_c\langle \sigma_{i-}\sigma_{i+}\rangle$ and cooling events generated by $\sigma_{i-}$ with probability $p(t)\Gamma_c(n_c+1)\langle\sigma_{i+}\sigma_{i-}\rangle$ acting on the ancilla qubits.

For simplicity we assume that the protocol takes place in 16 steps each of duration $\tau$.  The gates are carried out in sequence by setting the fields $f_i(t)$, $g_i(t)$, $h_i(t)$, and $k_i(t)$ to appropriate values, typically $\pi/2\tau$.  Note that the performance of the error correcting code depends directly on $\tau$ since the probability for a bit-flip error is $\gamma_h\tau$ per qubit.

\subsection{Numerical results}
\label{sec:results}
\begin{figure}
\center{\includegraphics[width=3in]{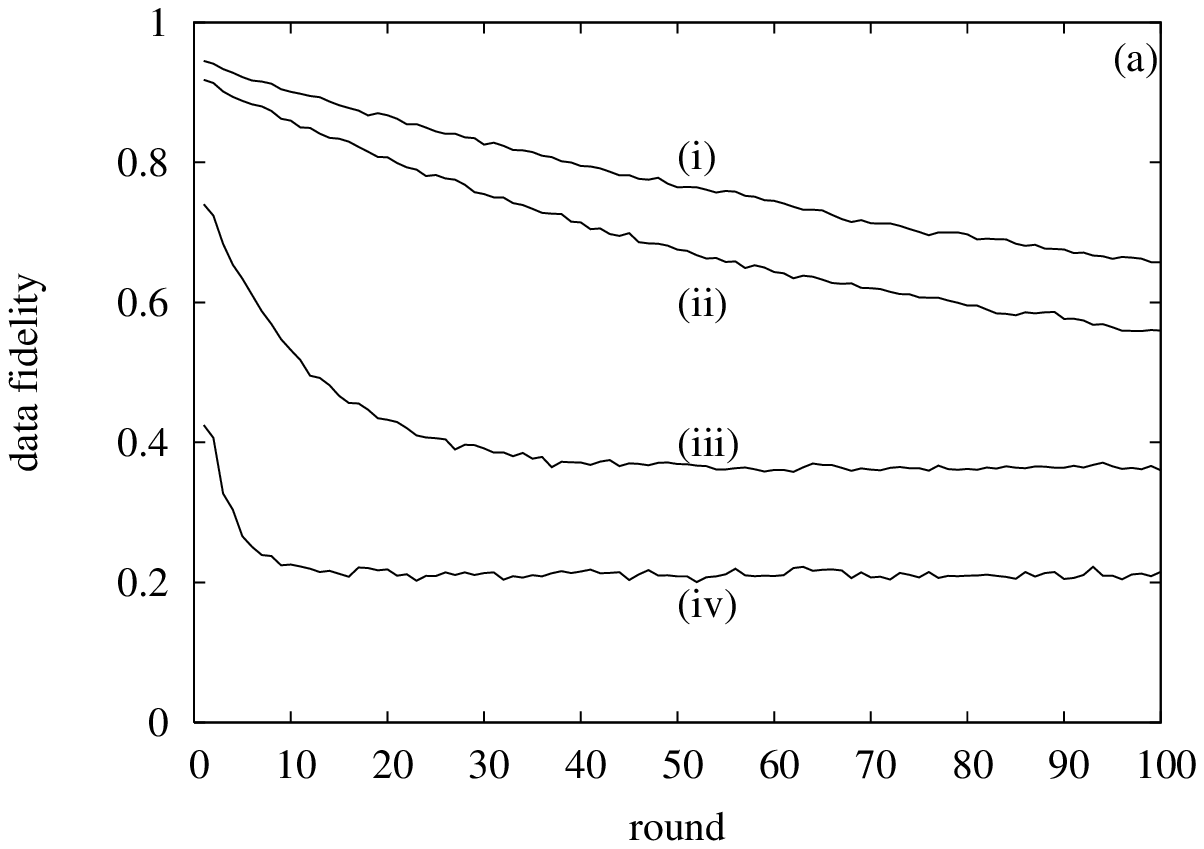}\includegraphics[width=3in]{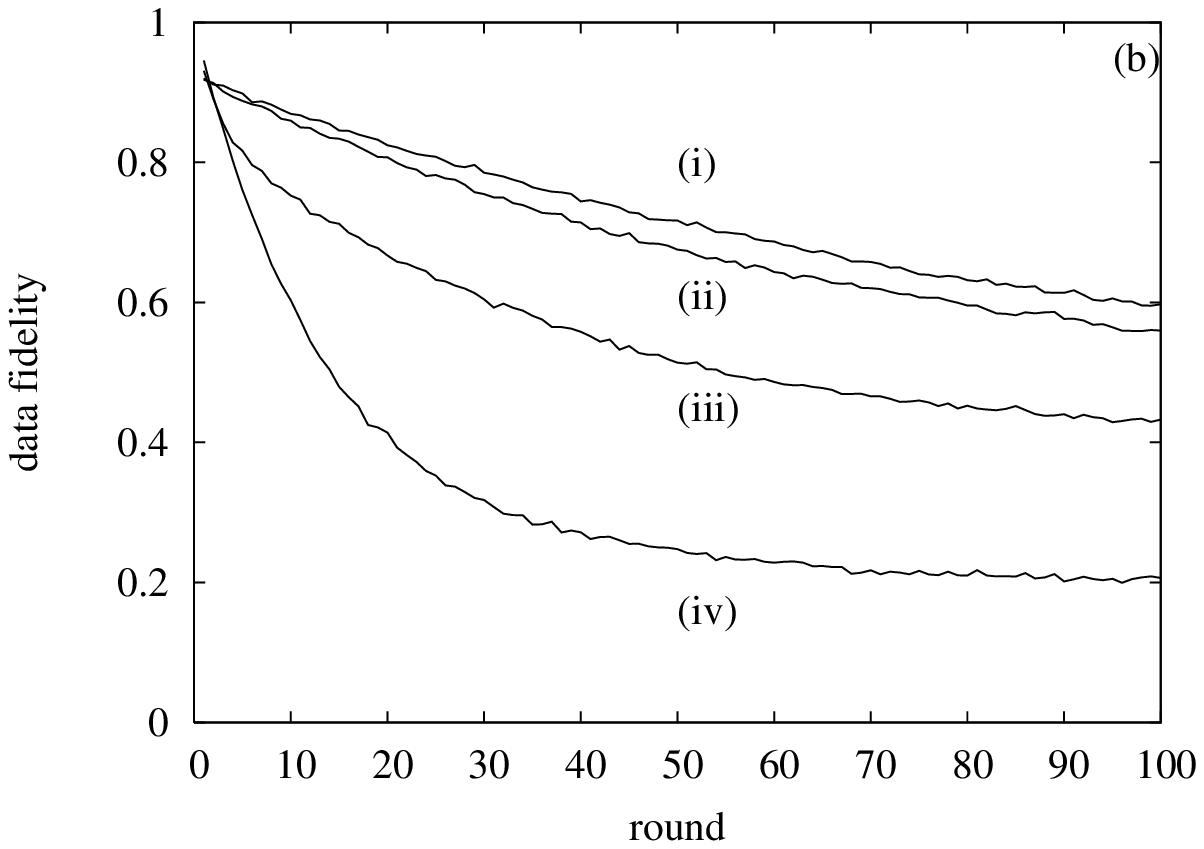}}
\center{\includegraphics[width=3in]{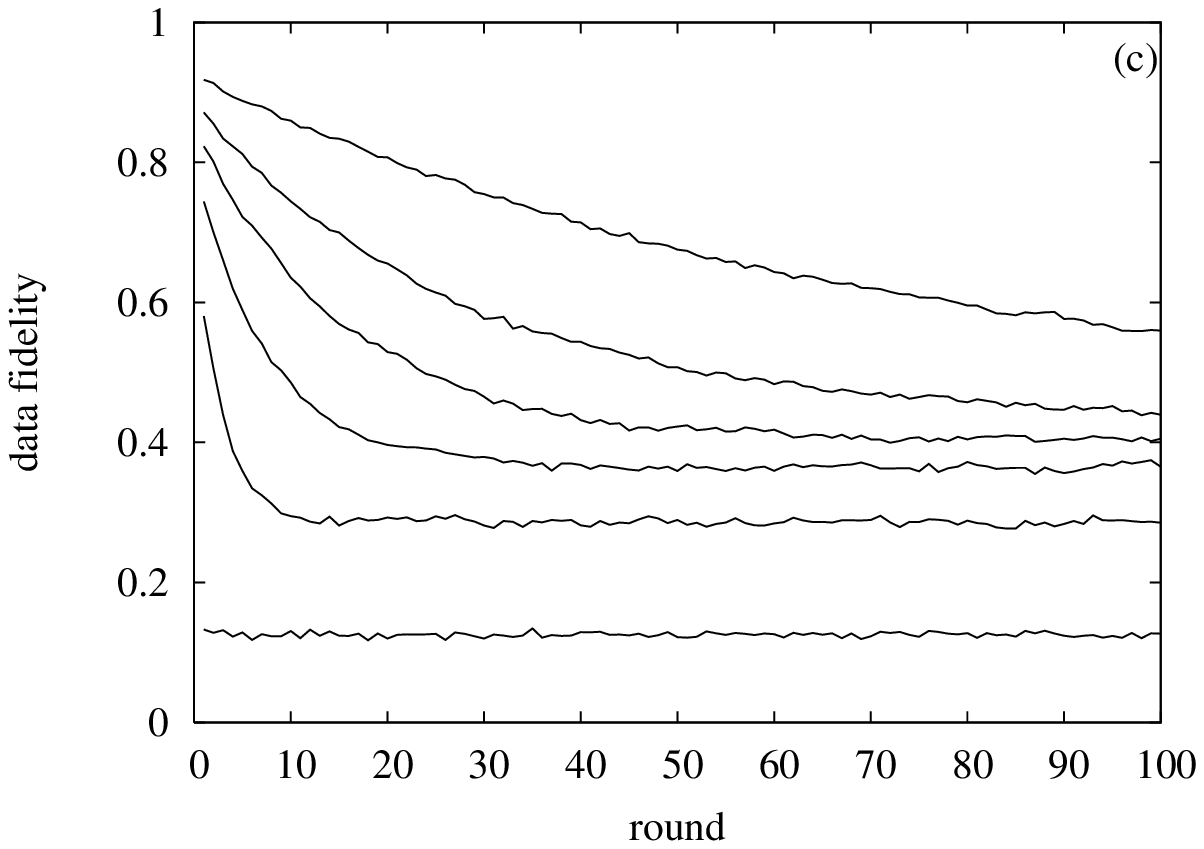}\includegraphics[width=3in]{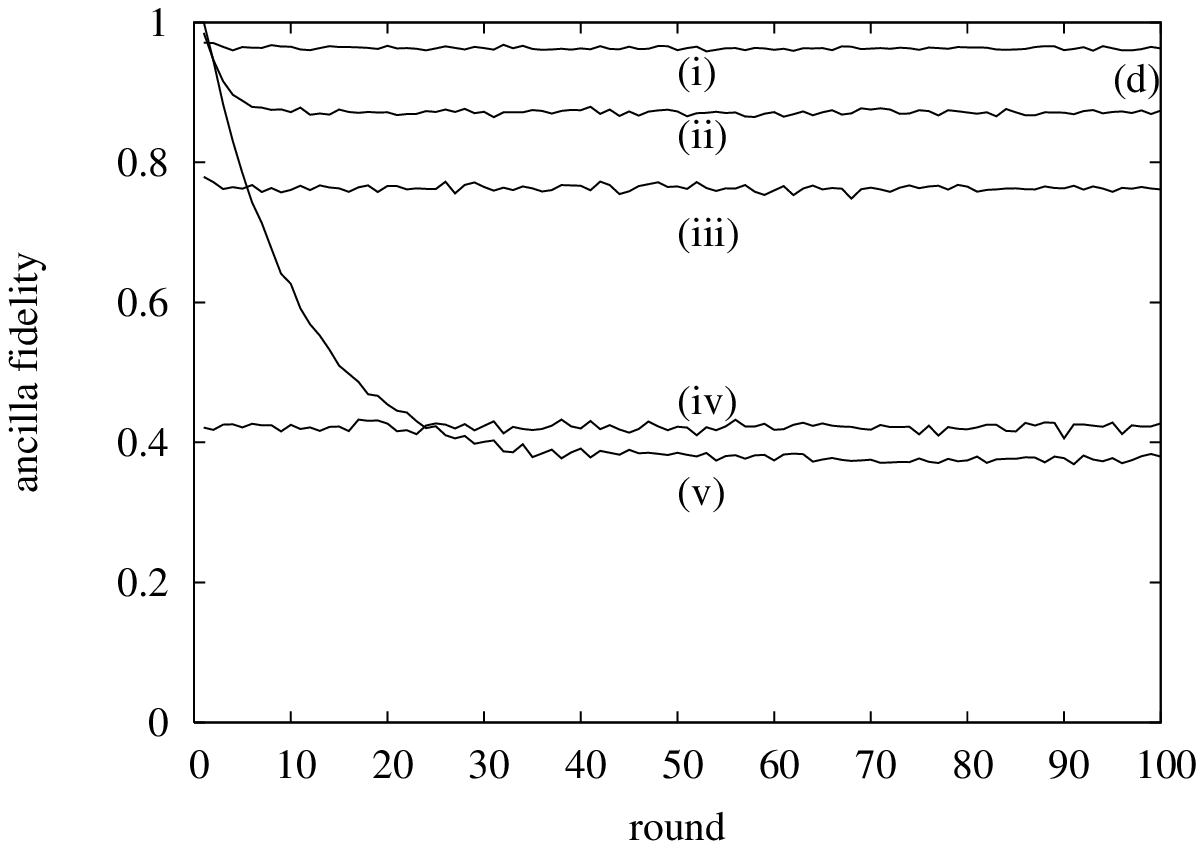}}
\caption{(a) Data fidelity for $\gamma_h=10^{-3}$, $\Gamma_c=3$ and $n_c = 0$ (i), $10^{-3}$ (ii), $10^{-2}$ (iii), $10^{-1}$ (iv). (b) Data fidelity for $n_c = 10^{-2}$, $\gamma_h = 10^{-3}$, and $\Gamma_c = 30$ (i), 3 (ii), 1 (iii), $10^{-1}$ (iv). (c) Data fidelity for $n_c = 10^{-2}$, $\Gamma_c = 3$, and $\gamma_h = 10^{-3} \times 1$, 2, 3, 5, 10, 100 from top curve to bottom curve. (d) Ancilla fidelity for $\gamma_h = 10^{-3}$ and $[n_c, \Gamma_c] = [10^{-2}, 3]$ (i), $[10^{-2},1]$ (ii), $[10^{-1}, 3]$ (iii), $[0.5, 3]$ (iv), $[10^{-2}, 10^{-1}]$ (v).\label{fig:fid-round}}
\end{figure}

First we consider the data fidelity following each round of error correction.  This is plotted in Fig.~\ref{fig:fid-round}(a) and (b) for $|\psi(0)\rangle = |0_L\rangle$.  In all cases the fidelity decays in time with the rate and steady state value depending on the quality of the ancilla preparation.  The heating rate is $10^{-3}$ for all curves in the plot.  One might at first expect the steady state value of the fidelity to be $1/8$ for the 3-bit code because there are $2^3 = 8$ states in the Hilbert space of the physical data qubits.  This assumes that the data qubits will explore all of the basis states with equal probability.  However, the error correction protocol drives the data qubits preferentially to the states $|000\rangle$ and $|111\rangle$ even if many errors have accumulated in the data qubits.  For very strong error correction the data qubits will effectively switch between $|000\rangle$ and $|111\rangle$ giving $F^2 = 1/2$ while for weak error correction the data qubits will explore all the basis states equally giving $F^2 = 1/8$.  Here strong error correction means that the protocol is likely to proceed without error for any single round while weak error correction means it is unlikely to proceed without error.


The effect of the heating rate on the data fidelity is plotted in Fig.~\ref{fig:fid-round}(c).  We have chosen values for $\gamma_c$ and $n_c$ which give reasonable ancilla cooling.  As the heating rate is increased the errors accumulate in the data qubits more rapidly while at the same time the ancilla qubits are less likely to be properly prepared before they are coupled to the data qubits..  For very strong heating, $\gamma_h = 0.1$, the fidelity is essentially constant at 1/8.


We also consider the ancillla fidelity immediately following the cooling step.  Several examples are plotted in Fig.~\ref{fig:fid-round}(d).  These curves split into two groups.  In the first (i, iii, iv) the ancilla fidelity is independent of the number of rounds preceding.  This is the case whenever the cooling rate $\Gamma_c$ is large enough ($\ge 3$) that effective cooling takes place.  Then the value of the ancilla fidelity is determined by $n_c$ or alternatively the temperature $T_c$.  The second group of curves (ii and v) show a decay of the fidelity in time to a steady state value.  This occurs when the cooling rate is too small.  In this case the cooling is not effective and the fidelity decays as errors accumulate in the ancilla qubits, either directly or by coupling to the data qubits.



\subsection{Ancilla fidelity}
We have been able to understand some of these behaviors with simple models.  For the ancilla cooling there are two types of extreme behavior to consider --- fast (lines i, iii, and iv in Fig.~\ref{fig:fid-round}d) and slow (line v in Fig.~\ref{fig:fid-round}d) cooling.  For fast cooling we assume the ancilla always reaches it's steady state during the cooling step.  The populations under cooling follow a set of rate equations
\begin{subequations}
\begin{eqnarray}
\dot P_0 &=& AP_1 + AP_2 + AP_4 - 3BP_0 \\
\dot P_1 &=& AP_3 + AP_5 - (A+2B)P_1 + BP_0 \\
\dot P_2 &=& AP_3 + AP_6 - (A+2B)P_2 + BP_0 \\
\dot P_3 &=& AP_7 - (2A+B)P_3 + BP_1 + BP_2 \\
\dot P_4 &=& AP_5 + AP_6 - (A+2B)P_4 + BP_0 \\
\dot P_5 &=& AP_7 - (2A+B)P_5 + BP_4 + BP_1 \\
\dot P_6 &=& AP_7 - (2A+B)P_6 + BP_4 + BP_2 \\
\dot P_7 &=& -3AP_7 + BP_6 + BP_5 + BP_3
\end{eqnarray}
\end{subequations}
where $P_i$ is the population of the $i$th ancilla state expressed as a decimal, $A = \Gamma_c(n_c+1)$, and $B = \Gamma_c n_c$.  In the steady state there is a symmetry $P_1 = P_2 = P_4 \equiv P_a$ and $P_3 = P_5 = P_6 \equiv P_b$ which reduces these equations to the four
\begin{subequations}
\begin{eqnarray}
0 &=& 3AP_a - 3BP_0 \\
0 &=& 2AP_b - (A+2B)P_a + BP_0 \\
0 &=& AP_7 - (2A+B)P_b + 2BP_a \\
0 &=& -3AP_7 + 3BP_b
\end{eqnarray}
\end{subequations}
with the normalization condition
\begin{equation}
P_0 + 3P_a + 3P_b + P_7 = 1.
\end{equation}
The steady state solution is
\begin{subequations}
\begin{eqnarray}
P_0 &=& (\frac{B}{A}+1)^{-3} \\
P_a &=& \frac{B}{A} P_0 \\
P_b &=& \left(\frac{B}{A}\right)^2 P_0 \\
P_7 &=& \left(\frac{B}{A}\right)^3 P_0
\end{eqnarray}
\end{subequations}
and the ancilla fidelity is
\begin{equation}
F^2 = P_0 = \left(\frac{n_c+1}{2n_c+1}\right)^3.
\end{equation}
Note that the deviation from perfect fidelity is due solely to the nonzero temperature of the cold reservoir.  Comparison with the simulation results shows good agreement.

Now consider the case of slow cooling.  If there is no cooling at all then the ancilla fidelity will decay because of heating of the ancilla qubits and the transfer of errors from the data qubits.  At each measurement step the ancilla qubits are projected onto a single basis state.  When there is effectively no cooling the fidelity is simply the probability that the ancilla is projected onto the $|000\rangle$ state.  This is determined in turn by the fidelity of the data qubits.  A comparison of the ancilla and data fidelities shows a close correlation for short times.  For long times the ancilla qubits reach a different steady state than the data qubits.

The steady state value of the ancilla qubits can be determined by a self-consistency condition from the dynamical equations governing the cooling.  These equations are the same as given above.  For simplicity we take $B=0$ and solve the equations for the individual initial conditions $P_i=1, P_{j\ne i}=0$ since those are appropriate following the measurement of the ancilla state.  The solutions are $i=0$:
\begin{equation}
P_0(t) = 1,
\end{equation}
$i=1, 2, 4$:
\begin{subequations}
\begin{eqnarray}
P_i(t) &=& e^{-At} \\
P_0(t) &=& 1-e^{-At},
\end{eqnarray}
\end{subequations}
$i=3,5,6$:
\begin{subequations}
\begin{eqnarray}
P_i(t) &=& e^{-2At} \\
P_{j,k}(t) &=& e^{At} - e^{-2At} \\
P_0(t) &=& 1-2e^{-At} + e^{-2At},
\end{eqnarray}
\end{subequations}
where $\{i,j,k\} = \{3,2,1\}, \{5,4,1\}, \{6,4,2\}$, and $i=7$:
\begin{subequations}
\begin{eqnarray}
P_7(t) &=& e^{-3At} \\
P_{3,5,6}(t) &=& e^{-2At} - e^{-3At} \\
P_{1,2,4}(t) &=& e^{-At} - 2e^{-2At} + e^{-3At} \\
P_0(t) &=& 1-3^{-At} + 3e^{-2At} - e^{-3At}.
\end{eqnarray}
\end{subequations}
Defining $x = e^{-At}$ we can write the fidelity as
\begin{equation}
F = p_0 + p_1(1-x) + p_2(1-x)^2 + p_3(1-x)^2
\end{equation}
where $p_i$ is the probability to have $i$ ancilla bits in state $|1\rangle$ prior to cooling.  At steady state these are related to the fidelity.  Roughly we can take $p_1 = (1-F_{ss}) + F_{ss}\alpha$ where $\alpha = 16 \,{\rm steps} \times \gamma_h \times 6\, {\rm qubits} = 96\gamma_h$ is the probability for the ancilla qubits to acquire an error during the round.  Note that errors on the data qubits are mapped onto the ancilla qubits and contribute to the ancilla infidelity in the limit of no cooling.  Putting $F_{ss}$ on both sides yields
\begin{equation}
F_{ss} = F_{ss}(1-\alpha) + \left[(1-F_{ss})(1-\alpha) + F_{ss}\alpha\right](1-x)
\end{equation}
and solving for $F_{ss}$ yields
\begin{equation}
F_{ss} = \frac{(1-\alpha)(1-x)}{1-\alpha-x+2\alpha x}.
\end{equation}
Again, comparison with the simulation results shows good agreement.

\subsection{Data fidelity}
The data fidelity following error correction can be found with a similar set of rate equations, however the behavior is more complicated than the cooling of the ancilla.  We trace the evolution of a set of probabilities $P_i$ from one round of error correction to the next.  There are 16 different events which can occur in one round of error correction.  The ancilla can be properly cooled (1) or not (2), the ancilla can be properly prepared (1) or not (2), and we can have 0 (1), 1 (2), 2 (3), or 3 (4) errors in the data qubits.  We label these events 111, 112, . . ., 224 accordingly.  The probabilities for these events to occur are
\begin{subequations}
\begin{eqnarray}
p_{111} &=& F_a\left[(1-\alpha)^3 + \alpha^3\right](1-\beta)^3 \\ 
p_{112} &=& F_a\left[(1-\alpha)^3 + \alpha^3\right] 3\beta(1-\beta)^2 \\
p_{113} &=& F_a\left[(1-\alpha)^3 + \alpha^3\right] 3\beta^2(1-\beta) \\
p_{114} &=& F_a\left[(1-\alpha)^3 + \alpha^3\right] \beta^3 \\
p_{121} &=& F_a\left[3\alpha(1-\alpha)^2 + 3\alpha^2(1-\alpha)\right] (1-\beta)^3 \\
p_{122} &=& F_a\left[3\alpha(1-\alpha)^2 + 3\alpha^2(1-\alpha)\right] 3\beta(1-\beta)^2 \\
p_{123} &=& F_a\left[3\alpha(1-\alpha)^2 + 3\alpha^2(1-\alpha)\right] 3\beta^2(1-\beta) \\
p_{124} &=& F_a\left[3\alpha(1-\alpha)^2 + 3\alpha^2(1-\alpha)\right] \beta^3 \\
p_{211} &=& p_{221} = (1-F_a)(1-\beta)^3 \\
p_{212} &=& p_{222} = (1-F_a)3\beta(1-\beta)^2 \\
p_{213} &=& p_{223} = (1-F_a)3\beta^2(1-\beta) \\
p_{214} &=& p_{224} = (1-F_a)\beta^3
\end{eqnarray}
\end{subequations}
where $F_a$ is the ancilla fidelity following cooling, $\alpha = 15\gamma_h$ is the probability per qubit for a single error in the ancilla qubits during a round of error correction, and $\beta=16\gamma_h$ is the probability per qubit for an error in the data qubits during a round of error correction.  Now we consider the flow of probability for each event.  Given that there are 0 data qubits in state $|1\rangle$ we have
\begin{subequations}
\begin{eqnarray}
f_{00} &=& p_{111} + p_{112} + \frac{1}{3}p_{122} + \frac{1}{3}p_{212} \\
f_{0a} &=& p_{121} + \frac{2}{3}p_{123} + p_{211} + \frac{2}{3}p_{213} \\
f_{0b} &=& p_{113} + \frac{2}{3}p_{122} + p_{124} + \frac{2}{3}p_{212} + p_{214} \\
f_{07} &=& p_{114} + \frac{1}{3}p_{123} + \frac{1}{3}p_{213}.
\end{eqnarray}
\end{subequations}
Given 1 data qubit in state $|1\rangle$ we have
\begin{subequations}
\begin{eqnarray}
f_{a0} &=& p_{111} + \frac{1}{3}p_{112} + \frac{1}{3}p_{121} + \frac{2}{9}p_{123} + \frac{1}{3}p_{211} + \frac{2}{3}p_{213} \\
f_{aa} &=& \frac{2}{3}p_{113} + \frac{7}{9}p_{122} + \frac{2}{3}p_{124} + \frac{7}{9}p_{212} + \frac{2}{3}p_{214} \\
f_{ab} &=& \frac{2}{3}p_{112} + p_{114} + \frac{2}{3}p_{121} + \frac{7}{9}p_{123} + \frac{2}{3}p_{211} + \frac{7}{9}p_{213} \\
f_{a7} &=& \frac{1}{3}p_{113} + \frac{2}{9}p_{122} + \frac{1}{3}p_{124} + \frac{2}{9}p_{212} + \frac{1}{3}p_{214}.
\end{eqnarray}
\end{subequations}
Given 2 or 3 data qubits we have the same equations as for 1 or 0 respectively with $P_0 \to P_7$, $P_a \to P_b$, $P_b \to P_a$, and $P_7 \to P_0$.  That is
\begin{subequations}
\begin{eqnarray}
f_{b0} &=& f_{a7}, \\
f_{ba} &=& f_{ab}, \\
f_{bb} &=& f_{aa}, \\
f_{b7} &=& f_{a0}, \\
f_{70} &=& f_{07}, \\
f_{7a} &=& f_{0b}, \\
f_{7b} &=& f_{0a}, \\
f_{77} &=& f_{00}.
\end{eqnarray}
\end{subequations}
This comes from the fact that the error correcting code treats $|000\rangle$ as equivalent to $|111\rangle$.

\begin{figure}
   \centering
   \includegraphics{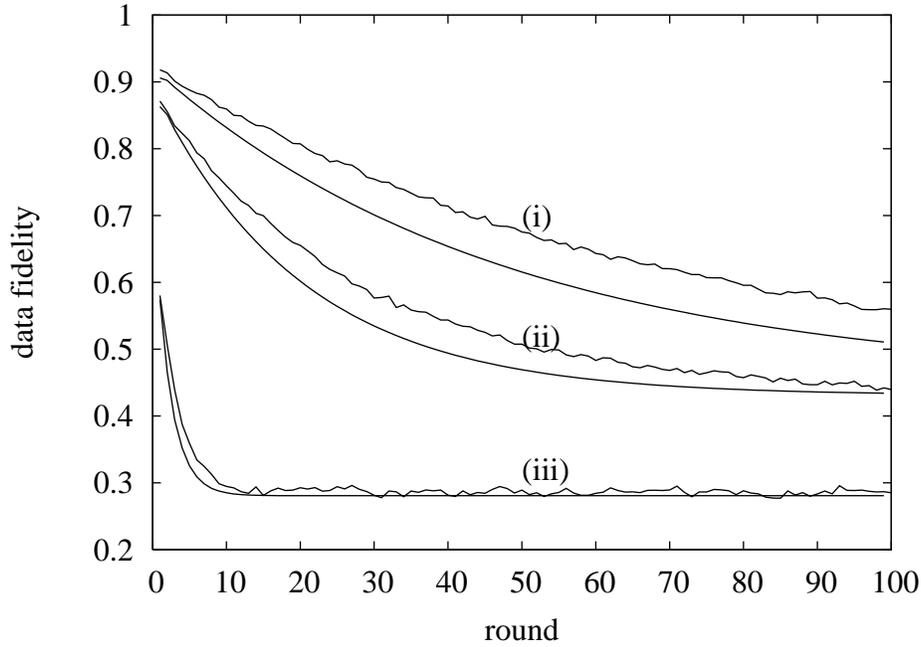} 
   \caption{Comparison of trajectory results and rate equations for $n_c = 10^{-2}$, $\Gamma_c = 3$, and $\gamma_h = 0.001$ (i), $0.002$ (ii), $0.1$ (iii).}
   \label{fig:compare}
\end{figure}

These have been presented as conditional probabilities, but they allow us to calculate the flow of probability from round $i$ of error correction to round $i+1$ when they are coupled with the actual probabilities from round $i$.  This takes the form of a set of rate equations
\begin{subequations}
\begin{eqnarray}
P_0(i+1) &=& P_0(i)f_{00} + P_a(i)f_{a0} + P_b(i)f_{b0} + P_7(i)f_{70}, \label{eq:rateeqns}\\
P_a(i+1) &=& P_0(i)f_{0a} + P_a(i)f_{aa} + P_b(i)f_{ba} + P_7(i)f_{7a}, \\
P_b(i+1) &=& P_0(i)f_{0b} + P_a(i)f_{ab} + P_b(i)f_{bb} + P_7(i)f_{7b}, \\
P_7(i+1) &=& P_0(i)f_{07} + P_a(i)f_{a7} + P_b(i)f_{b7} + P_7(i)f_{77},
\end{eqnarray}
\end{subequations}
which we have solved numerically to produce plots in Fig.~\ref{fig:compare} of the data fidelity for comparison with the simulations.

It turns out that for the case of good ancilla cooling the decay of the fidelity is very nearly exponential following a relatively large decrease in the initial rounds of error correction.  We consider these two features in turn.  The large initial decay in the data fidelity can be found by considering $P_0(1)$ from equation \eqref{eq:rateeqns},
\begin{equation}
P_0(1) = f_{00} = \left(\frac{n_c+1}{2n_c+1}\right)^3\left(1-3\alpha-\beta\right) + \beta
\end{equation}
where we have used the definitions from Eqs. (10), (18), and (19) keeping only first order terms in $\alpha$ and $\beta$.  The form of the fidelity following one round of error correction reveals that the source of the large initial decay in the fidelity is a combination of non-zero temperature of the cold reservoir ($n_c \ne 0$), errors on the ancilla qubits, and errors on the data qubits.  Either imperfect ancilla cooling or errors on the ancilla qubits during preparation will always contribute to a reduced data fidelity following the first round of error correction while errors in the data qubits only contribute in combination with imperfect ancilla cooling.

We quantify the subsequent slow decay of the fidelity by solving  the rate equations to a few orders in the heating rate $\alpha$ and matching term by term with decaying exponentials of the form
\begin{subequations}
\begin{eqnarray}
P_0(n) &=& P_{0,\rm ss} + (P_0(k) - P_{0,\rm ss}) \delta_0^{-(n-k)} \label{eq:exp}\\
P_a(n) &=& P_{a,\rm ss} - (P_{a,\rm ss} - P_a(k)) \delta_a^{-(n-k)} \\
P_b(n) &=& P_{b,\rm ss} - (P_{b,\rm ss} - P_b(k))\delta_b^{-(n-k)} \\
P_7(n) &=& P_{7,\rm ss} - (P_{7,\rm ss} - P_7(k))\delta_7^{-(n-k)}
\end{eqnarray}
\end{subequations}
and find the form of $\delta$ to the same order in $\alpha$ where $k$ is a small number to account for the large initial decay of the fidelity.  Here we take $\beta = \alpha$ and $F_a = 1$ for simplicity.  To illustrate this, consider the fidelity written to second order in $\alpha$
\begin{eqnarray}
P_0(n) &=& f_{00}^n + (n-1)f_{00}^{n-2}f_{0a}f_{a0} + \frac{(n-2)(n-3)}{2}f_{0a}^2 \\
&=& 1-3\alpha + (33-21n)\alpha^2 \nonumber
\end{eqnarray}
with the steady state value
\begin{eqnarray}
P_{0,\rm ss} &=& \frac{1}{2}\frac{f_{10} + f_{13}}{f_{01} + f_{02} + f_{10} + f_{13}} \\
&=&  \frac{1}{2}\left(1-3\alpha+24\alpha^2\right). \nonumber
\end{eqnarray}
We write $\delta_0$ as
\begin{equation}
\delta_0 = a_0\left(1 + a_1\alpha + a_2\alpha^2\right)
\end{equation}
where $a_i$ are coefficients to be determined.  We choose $k=4$ in order to keep all the terms in $P_0(n)$ to second order and write out Eq.~\ref{eq:exp} as
\begin{eqnarray}
P_0(n) = \frac{1}{2} - \frac{3}{2}\alpha + 12\alpha^2 + a_0^{-(n-4)}\Bigg[\frac{1}{2} &+& \left(2a_1-\frac{3}{2}\right)\alpha + \left(2a_2+3a_1^2-6a_1-63\right)\alpha^2 - \frac{1}{2}a_1n\alpha \nonumber \\
&&+ \left(\frac{3}{2}a_1 - \frac{7}{4}a_1^2 - \frac{1}{2}a_2\right) n\alpha^2 + \frac{1}{4}a_1^2 n^2\alpha^2\Bigg] 
\end{eqnarray}
Matching the constant term requires $a_0 = 1$, matching the $\alpha$ term requires $a_1=0$, and matching the $n \alpha$ term requires $a_2 = 42$.  Then we have $\delta_0 = 1+42\alpha^2$.

\subsection{Cyclic features in error correction}
\begin{figure}
\center{\includegraphics[width=3in]{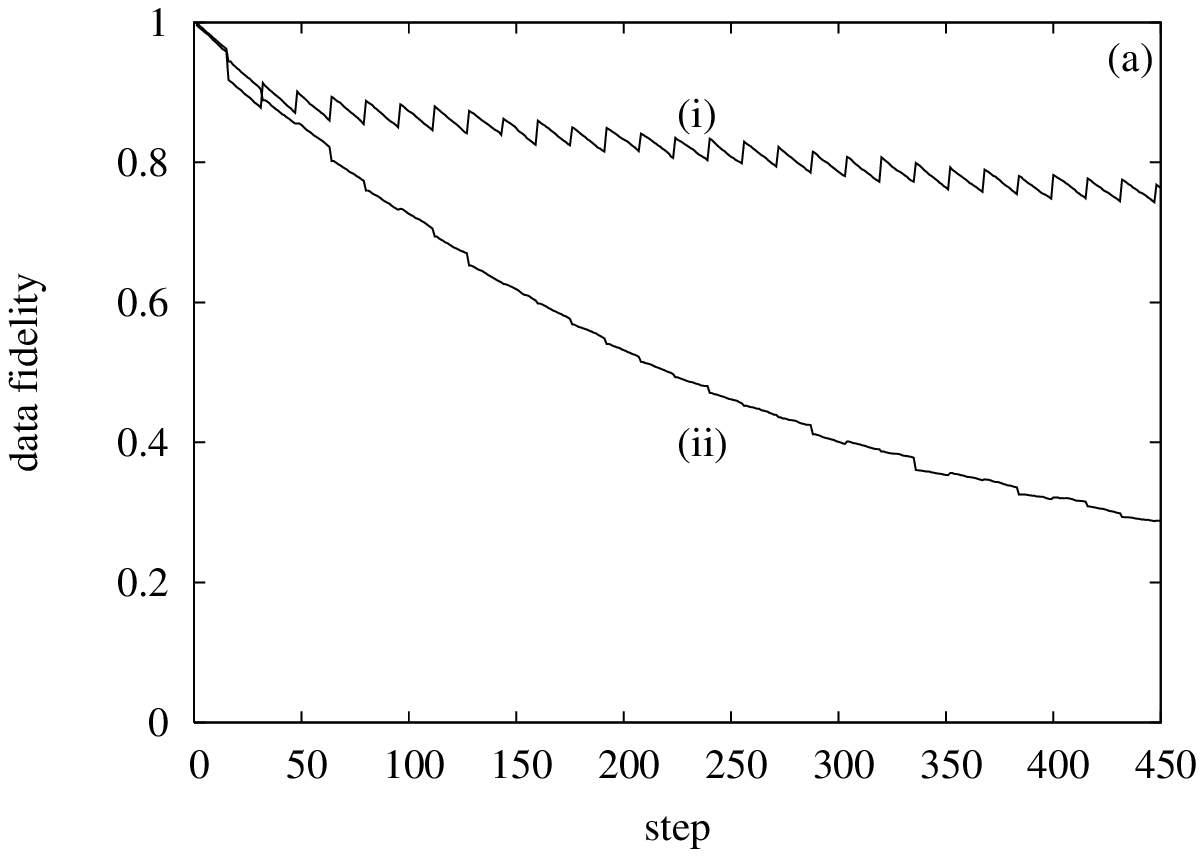}\includegraphics[width=3in]{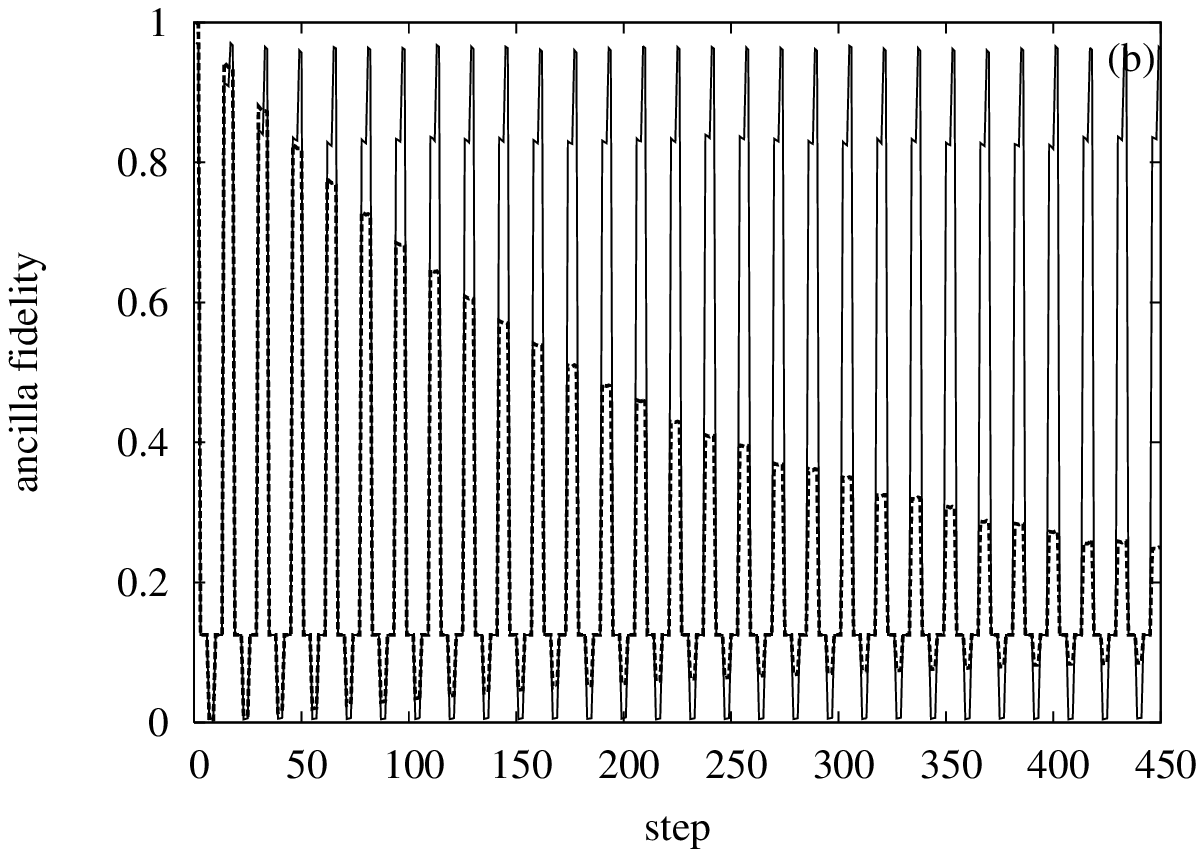}}
\center{\includegraphics[width=3in]{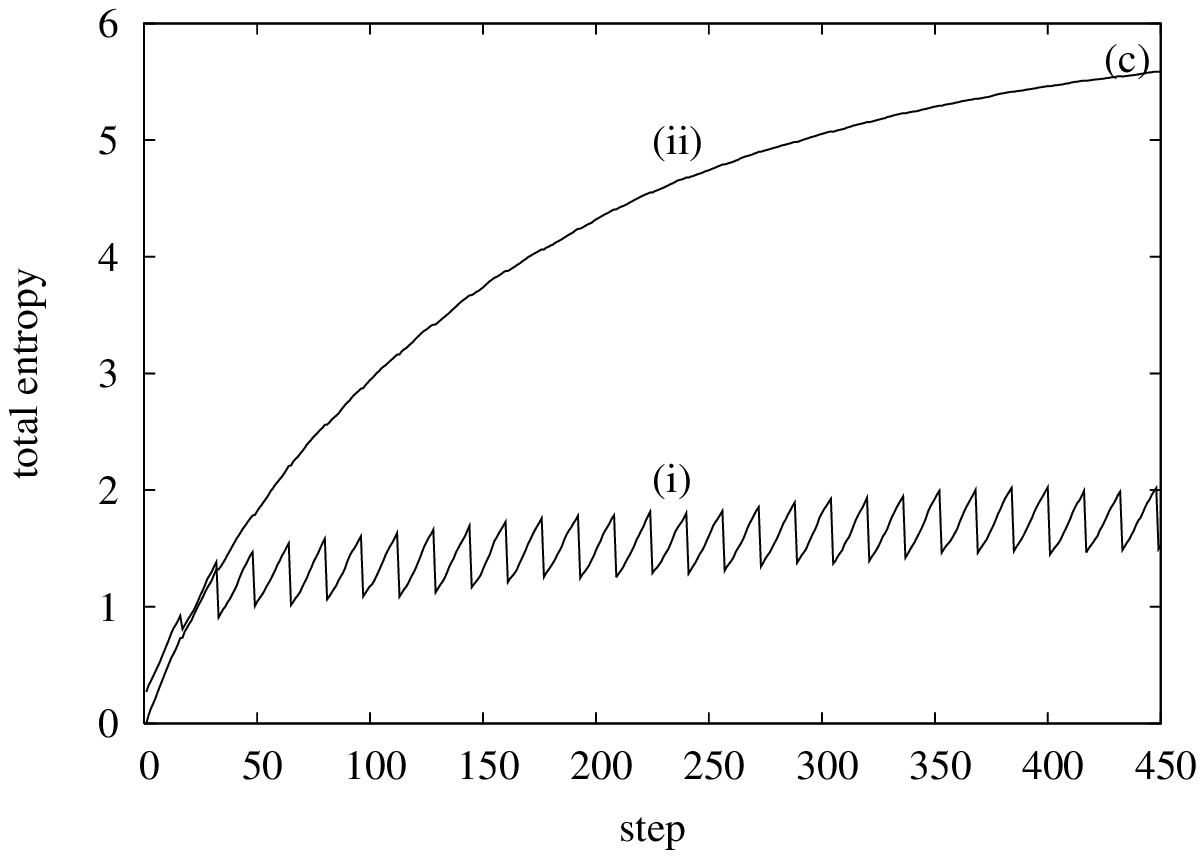}\includegraphics[width=3in]{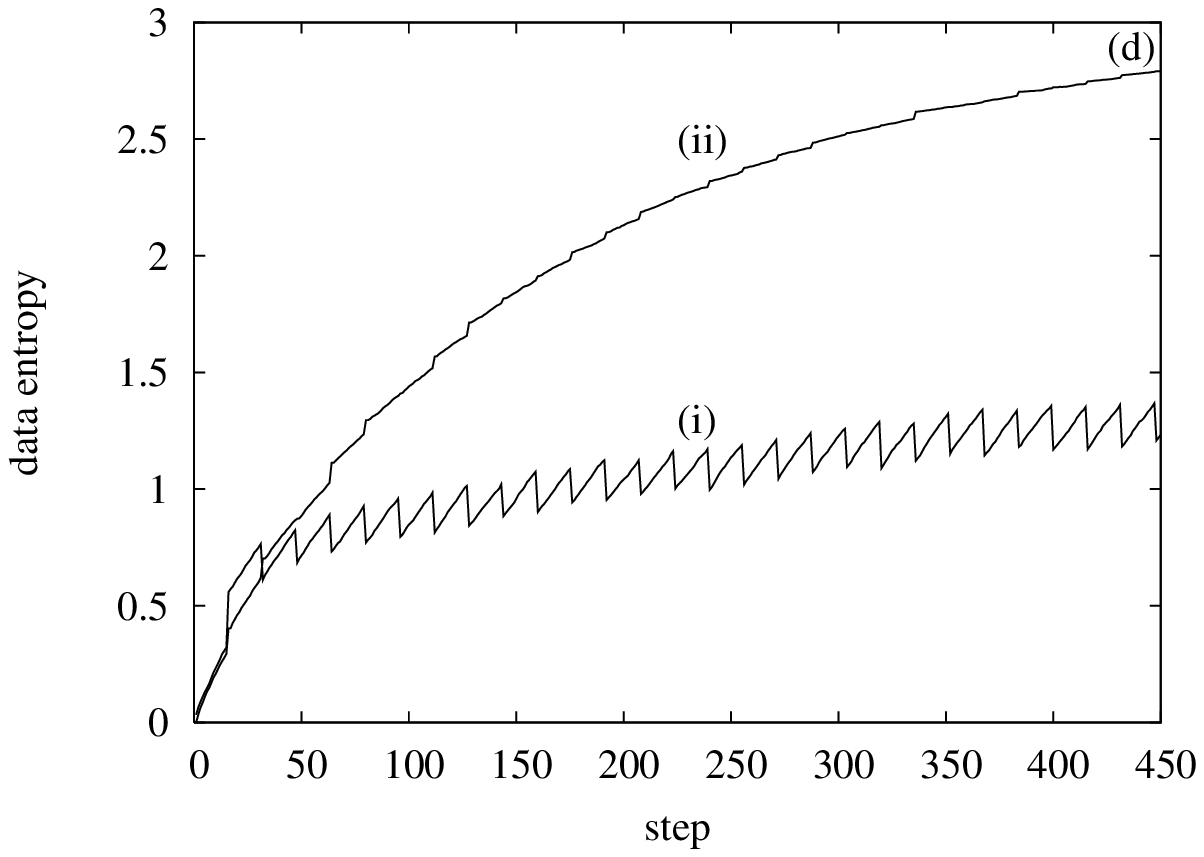}}
\center{\includegraphics[width=3in]{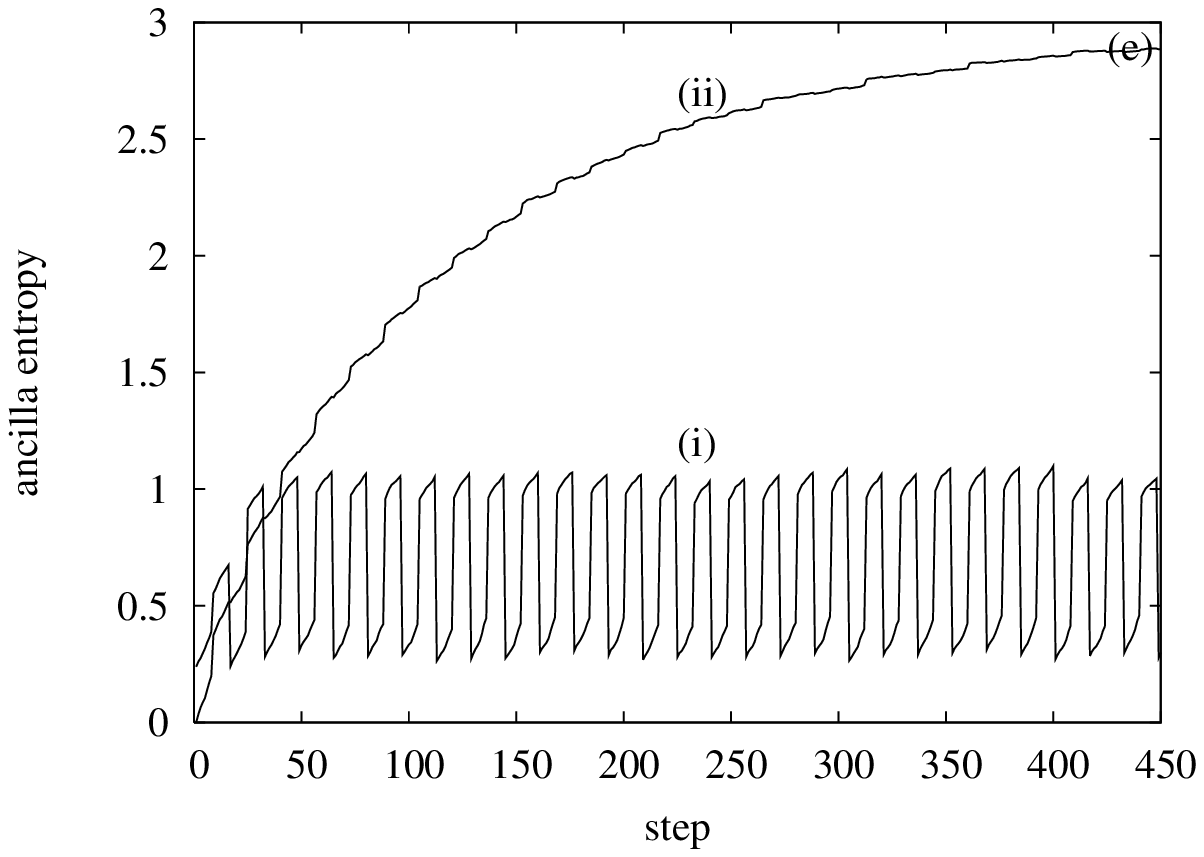}}
\caption{\label{fig:cycles}Data fidelity (a), ancilla fidelity (b), total entropy (c), data entropy (d), and ancilla entropy (e) for $[n_c, \Gamma_c, \gamma_h] = [10^{-2}, 3, 10^{-3}]$ (i) and $[10^{-2}, 10^{-2}, 10^{-3}]$ (ii).  In plot (b) the thin line corresponds to parameters (i) and the thick line corresponds to parameters (ii).}
\end{figure}

So far we have considered the gradual decay of the fidelity over many rounds of error correction.  Now we look at the fidelity and the entropy at each step of the error correction protocol (see Fig.~\ref{fig:expandprot}).  We use the von Neumann entropy defined as
\begin{equation}
S = -{\rm tr}\left(\rho\log\rho\right)
\end{equation}
where the density operator $\rho$ is found by averaging $|\psi_{\rm REC}(t)\rangle\langle\psi_{\rm REC}(t)|$ over many trajectories.  These are plotted in Fig.~\ref{fig:cycles} showing the cyclic process when the error correction is working, particularly for the ancilla qubits.  We have chosen $\gamma_h = 10^{-3}$ and two sets of parameters:  $\{n_c,\gamma_c\} = \{10^{-2}, 3\}$ (i) for effective error correction and $\{10^{-2}, 10^{-2}\}$ (ii) for ineffective error correction.  The difference is apparent in both the fidelity and the entropy.  When the error correction is effective there are periodic increases in the fidelity and corresponding decreases in the entropy which appear in the time step in which the correction is applied.  When the error correction is ineffective there is simply a steady decrease in the fidelity and increase in the entropy of the data qubits.

It is worthwhile considering the physical reasons for the differences between the sets of curves labeled (i) and (ii) in Fig.~\ref{fig:cycles}.  In the curves (i) the ancilla is well-cooled; after every error-correction cycle, its entropy goes down to the same ``floor'' value and the ancilla fidelity rises to the same ``ceiling''.  That this floor is not zero is due to the nonzero reservoir temperature represented by $n_c$ and the finite cooling rate $\Gamma_c$.  However, even though the ancilla entropy does not increase on average in this case, the data entropy continues to increase on average with small decreases upon successful error correction;  this is due to the fact that there is always a probability that an uncorrectable error will take place (an effect proportional to $\gamma_h^2$).  The nonzero ancilla entropy (and thus $n_c$) also plays a role in the steady increase of the data entropy because an incorrectly prepared ancilla will cause the wrong syndrome to be extracted so the data gets miscorrected.

For the curves labeled (ii) there is effectively no cooling of the ancilla qubits and the ancilla entropy steadily increases.  Without a properly prepared ancilla the syndrome is not extracted properly and the errors in the data qubits cannot be corrected.






\section{Error Correction Without Measurement\label{sec:nomeasure}}
\begin{figure}
\includegraphics{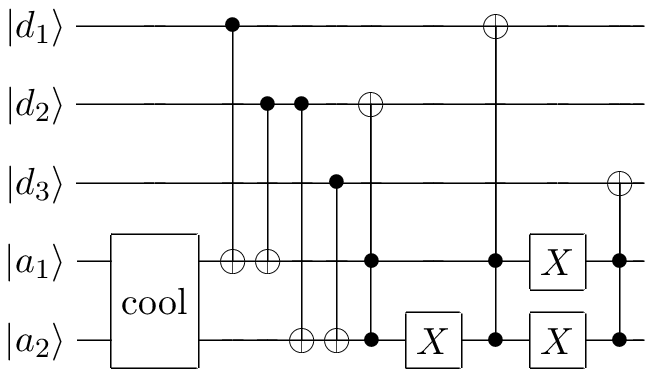}
\caption{\label{fig:nomeasure}The quantum circuit diagram for implementing the three qubit repetition code without measurement.}
\end{figure}

It is also possible to implement error correction without measurement by means of additional quantum gates to carry out the appropriate correction conditionally based on the state of the ancilla qubits.  A circuit for implementing the three bit repetition code using two ancilla qubits is shown in Fig.~\ref{fig:nomeasure}.  The correction is carried out by means of three Toffoli gates which flip the approriate data qubit based on the state of the ancilla qubits.

\begin{figure}
\includegraphics[width=6.5in]{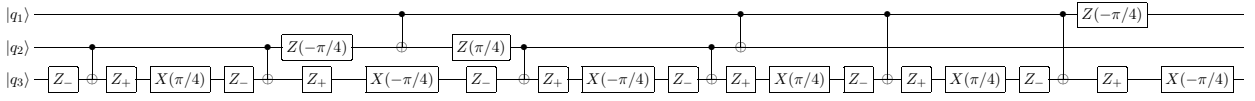}
\caption{\label{fig:expandtoffoli}The quantum circuit to implement a Toffoli gate using one- and two-bit gates.}
\end{figure}

We use the same Hamiltonian to simulate this quantum circuit with an appropriate sequence of control fields.  In order to do this we must expand each Toffoli gate into a sequence of one- and two-bit gates.  Based on results in Ref.~\cite{Barenco-1995-3457} one such circuit is shown in Fig.~\ref{fig:expandtoffoli}.  The CNOT gates appearing in the error correction circuit and the Toffoli gates are again expanded according to Fig.~\ref{fig:expandcnot} and a sequence of fundamental gates is found to implement the error correction without measurement in 68 equal time steps.

It should be noted that this scheme is quite similar to that described in Ref.~\cite{Sarovar-2005-012306} but with three important differences.  First, Sarovar and Milburn consider a continuous coupling of the ancilla qubits to a zero temperature reservoir while we assume that the ancilla qubits are cooled for a single time step at the beginning of the protocol by coupling to a finite temperature reservoir.  Second, we include heating of the ancilla qubits as well as the data qubits.  Third, they use a time independent Hamiltonian to carry out the error correction but it is not clear how that might be realized experimentally.  We have used an explicit sequence of gates with a time dependent Hamiltonian which could be implemented in a variety of physical systems.

\begin{figure}
\includegraphics[width=3in]{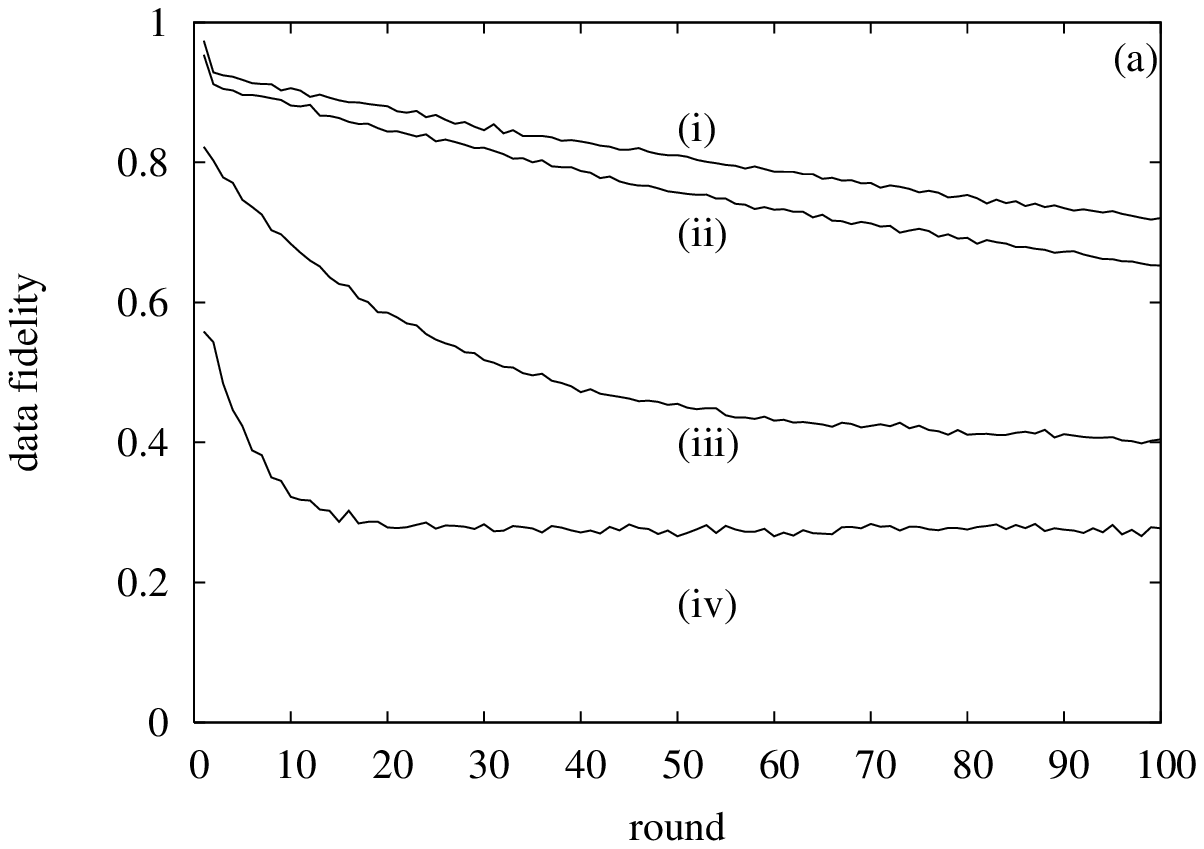}\includegraphics[width=3in]{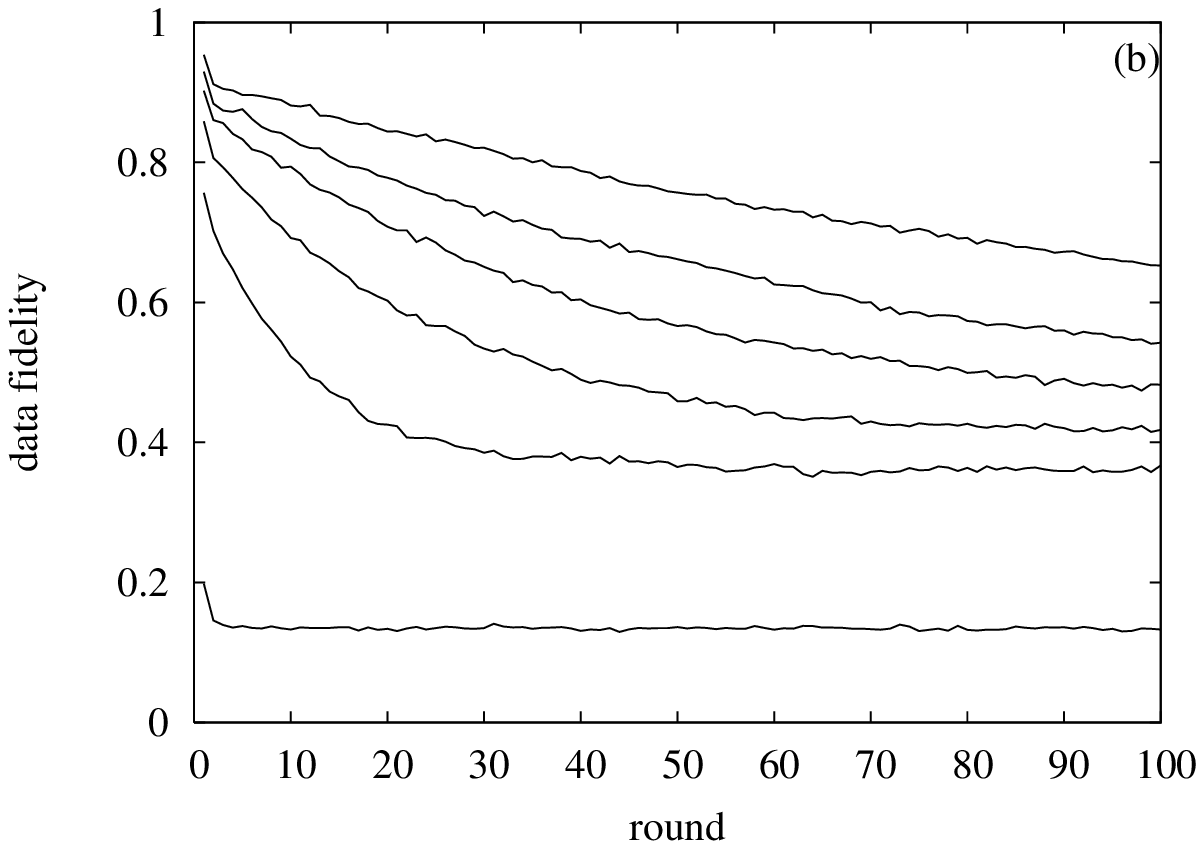}
\caption{\label{fig:round-nm}(a) Data fidelity for $\gamma_h=10^{-4}$, $\Gamma_c=3$ and $n_c = 0$ (i), $10^{-3}$ (ii), $10^{-2}$ (iii), $10^{-1}$ (iv).  (b) Data fidelity for $n_c = 10^{-2}$, $\Gamma_c = 3$, and $\gamma_h = 10^{-4} \times 1$, 2, 3, 5, 10, 100 from top curve to bottom curve.}
\end{figure}

First we consider the data fidelity following each round of error correction plotted in Fig.~\ref{fig:round-nm} for $|\psi(0)\rangle = |0_L\rangle$.  In all cases the fidelity decays in time with the rate and steady state value depending on the quality of the ancilla preparation.  The heating rate is $10^{-4}$ for all curves in the plot (a).  Just as for error correction with measurement, one might expect the steady state value of the fidelity to be $1/8$ for the 3-bit code because there are $2^3 = 8$ states in the Hilbert space of the physical data qubits.  However, for very strong error correction the data qubits will effectively switch between $|000\rangle$ and $|111\rangle$ giving $F^2 = 1/2$ while for weak error correction the data qubits will explore all the basis states equally giving $F^2 = 1/8$.

The effect of the heating rate on the data fidelity is plotted in Fig.~\ref{fig:fid-round}(b).  We have chosen values for $\gamma_c$ and $n_c$ which give reasonable ancilla cooling.  As the heating rate is increased the errors accumulate in the data qubits more rapidly while at the same time the ancilla qubits are less likely to be properly prepared.  For very strong heating, $\gamma_h = 0.01$, the fidelity is essentially constant at 1/8.  Notice that $\gamma_h$ is an order of magnitude smaller in these plots that for error correction with measurement shown in Fig.~\ref{fig:fid-round}.  This is necessary because it takes many time steps to carry out the Toffoli gates giving a larger failure rate per round of error correction.

\begin{figure}
\includegraphics[width=3in]{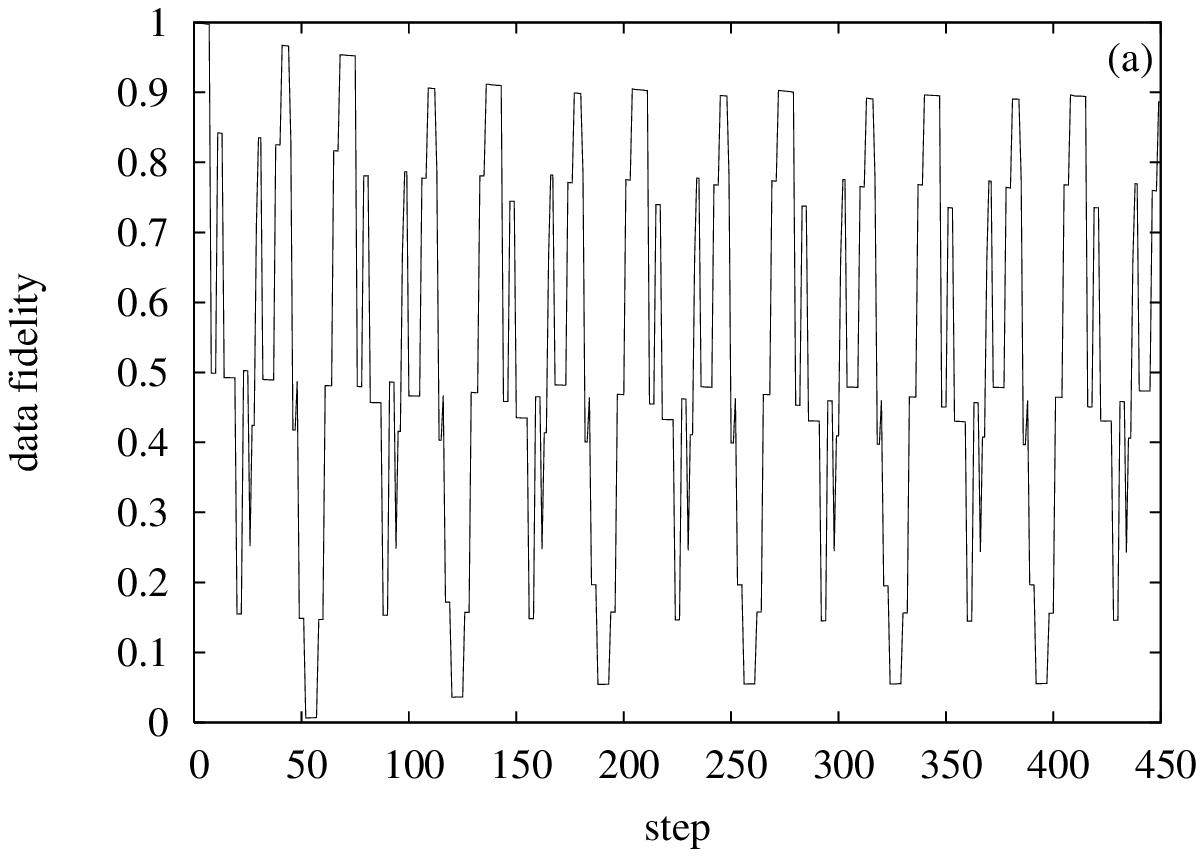}\includegraphics[width=3in]{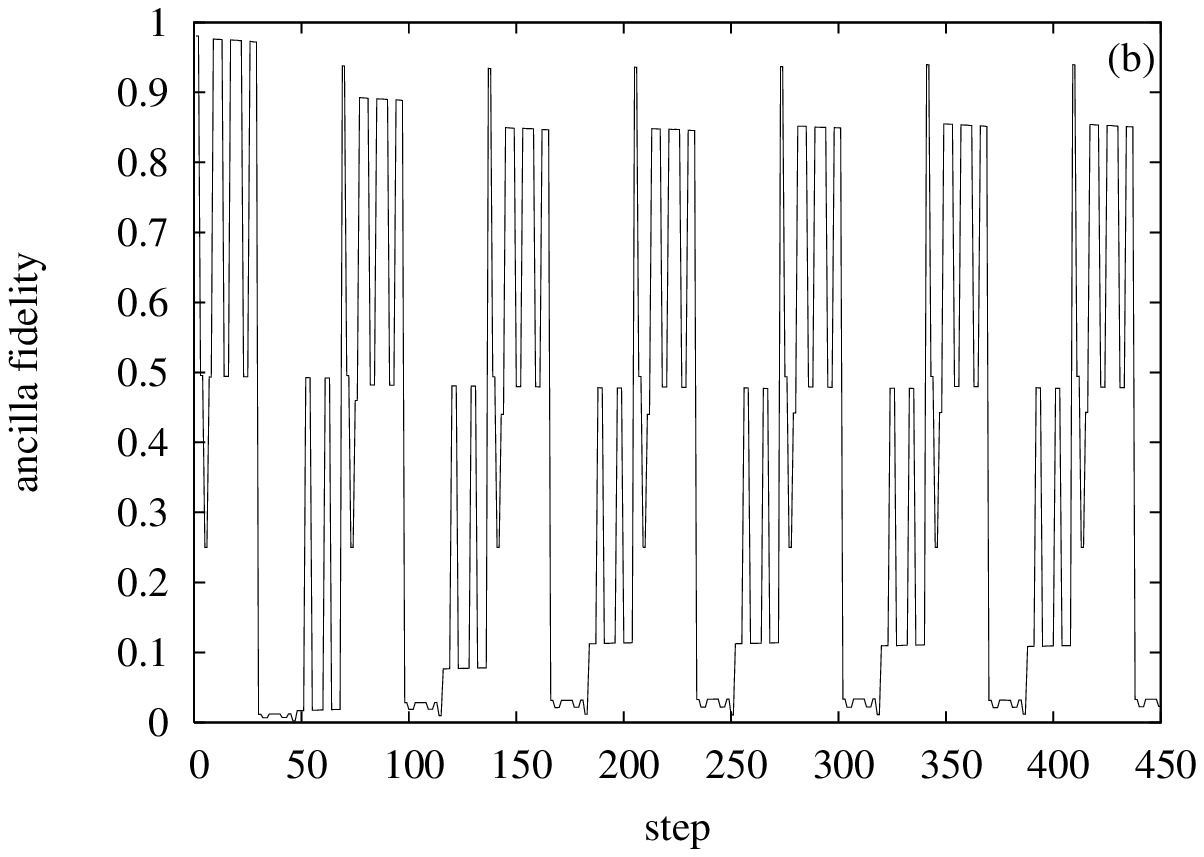}
\includegraphics[width=3in]{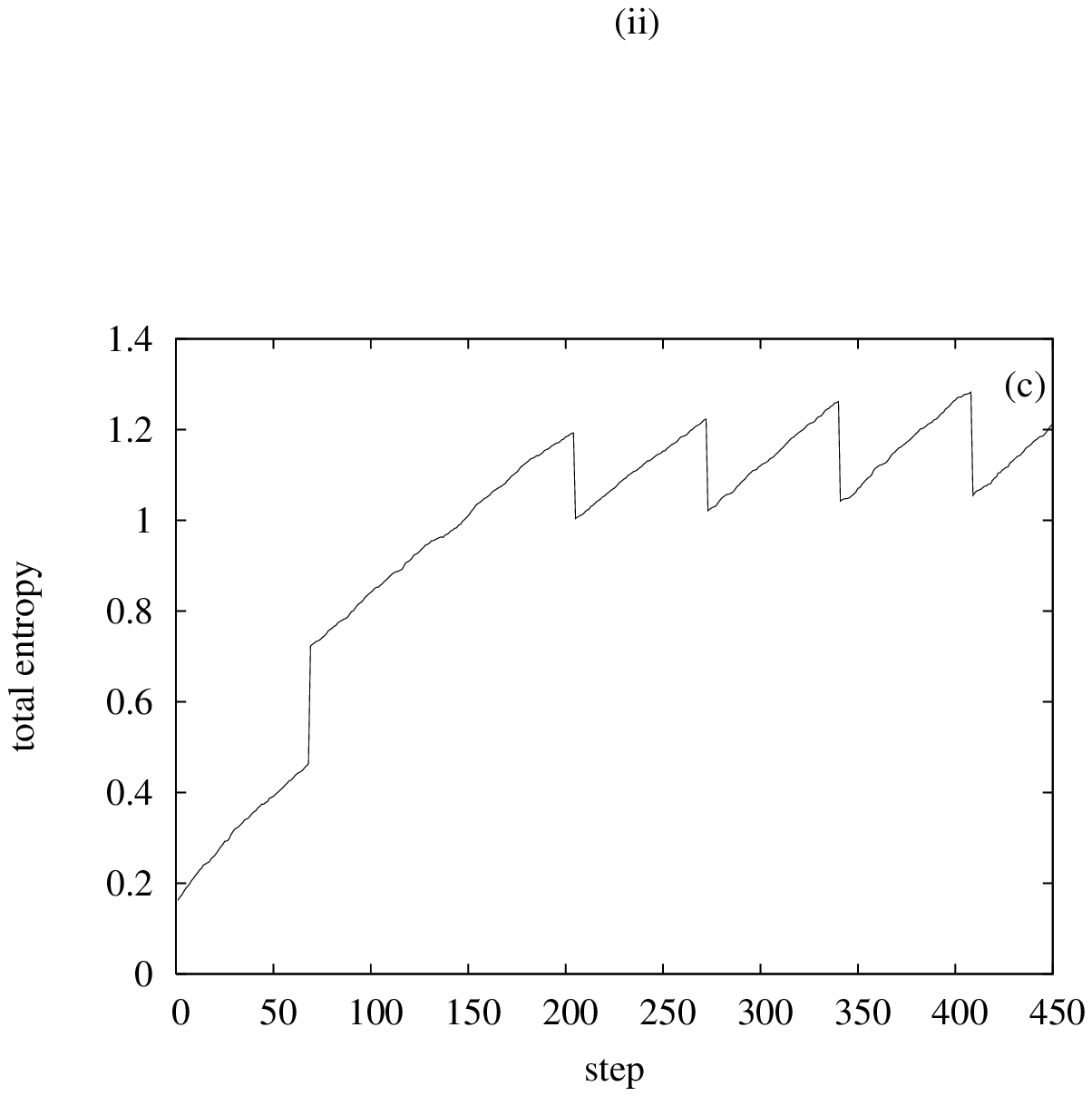}\includegraphics[width=3in]{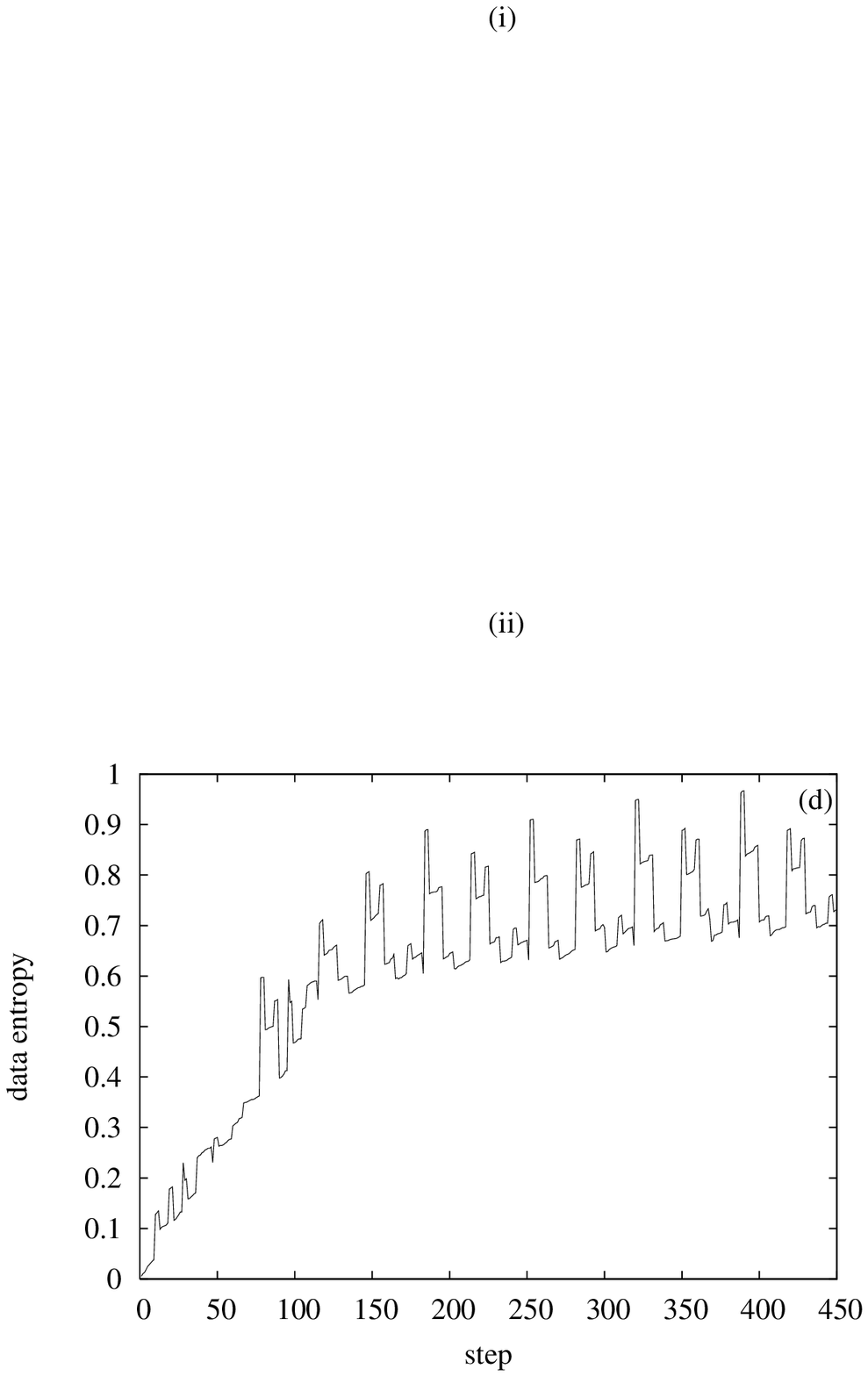}
\includegraphics[width=3in]{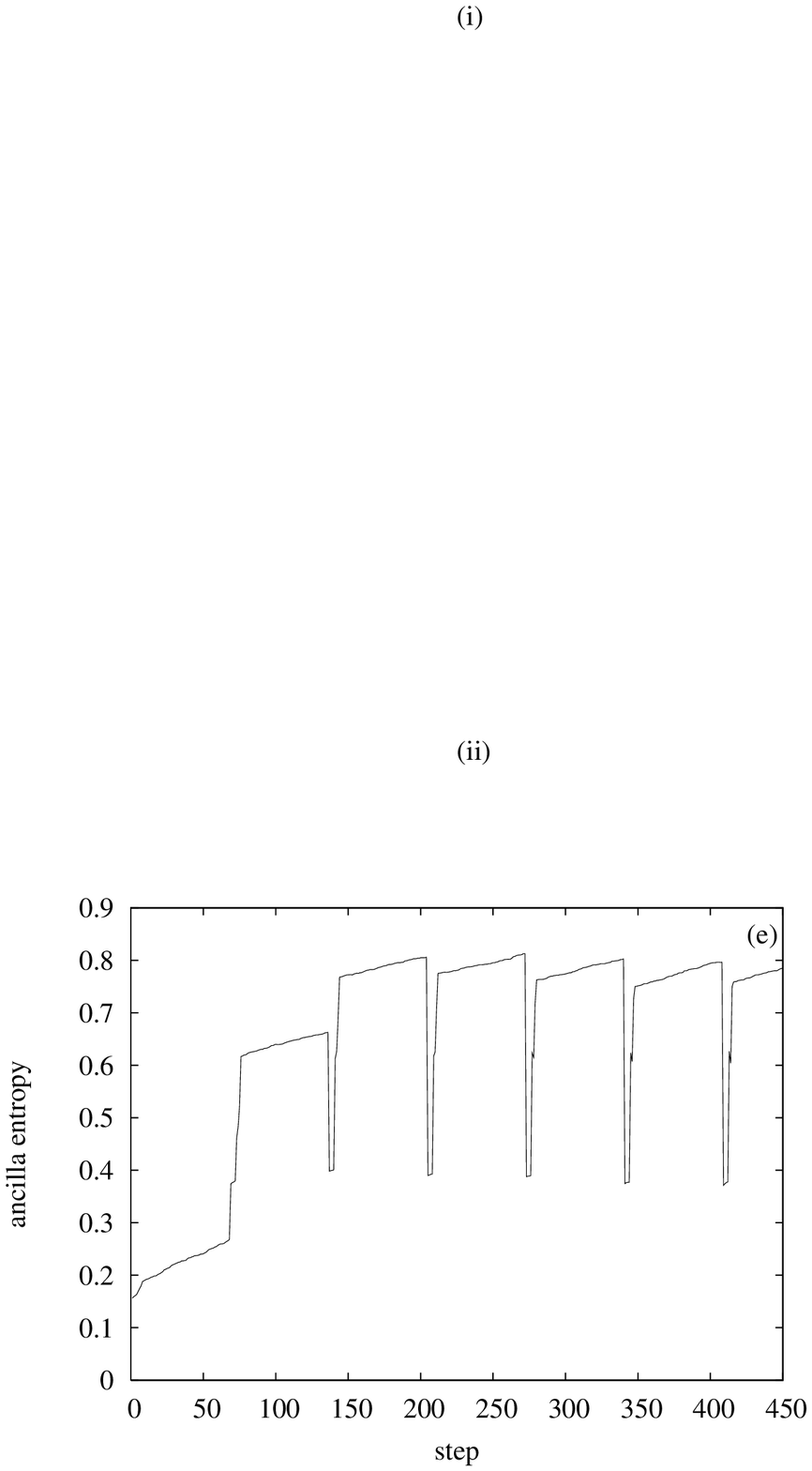}
\caption{\label{fig:cycles-nm}Data fidelity (a), ancilla fidelity (b), total entropy (c), data entropy (d), and ancilla entropy (e) for $[n_c, \Gamma_c, \gamma_h] = [10^{-2}, 3, 10^{-4}]$.}
\end{figure}

The fidelity and entropy for error correction without measurement is shown in Fig.~\ref{fig:cycles-nm} for each time step.  Here the ancilla qubits have a strong cyclic behavior while the data qubits show an overall increase in entropy along with cyclic variations.

\section{Discussion and conclusions}
\label{sec:discuss}
Our analysis has revealed a similar cyclic pumping of entropy from the data qubits to the ancilla qubits to the cold reservoir in for error correction with and without measurement.  They also share a similar decay of the fidelity of the data qubits due to incomplete removal of the entropy with a fast initial decay followed by a slower decay as the total system reaches a steady state.  The performance is quite sensitive to the ancilla cooling with either slow cooling or cooling to a nonzero temperature leading to degraded performance of the error correction protocol.  It is interesting to note that in these circumstances the decay of the fidelity is first order in the heating rate as a single bit flip in the ancilla qubits is enough to disrupt the error correction protocol.

In fault-tolerant quantum error correction schemes (see Ref.~\cite{Steane-2003-042322} for example) the linear dependence on the error rate is addressed in two ways.  The ancilla state is verified prior to coupling to the data qubits leading to a reduction in the probability that an incorrectly prepared ancilla state leads to an incorrect syndrome.  Furthermore, multiple syndrome extraction leads to a reduction in the probability that the error in the data qubits is incorrectly diagnosed and therefore miscorrected.

It should also be noted that the ancilla cooling imposes a minimum energy requirement (this is separate from the minimum energy requirements for quantum logic discussed by one of us in \cite{Gea-Banacloche-2002-217901}) to perform error correction due to Landauer's priniciple which requires $k_B T_c \ln 2$ of energy per bit of information erased.

We acknowledge support for this work from the U.~S.~Army Research Office.

\bibliography{QECCbib}

\end{document}